\documentclass[pra,showpacs,superscriptaddress]{revtex4-1}
\usepackage{epsf,epsfig}
\usepackage[psamsfonts]{amssymb}
\usepackage{bm}
\usepackage{graphicx}
\usepackage{dsfont}
\usepackage{amsmath}
\usepackage{natbib}
\usepackage{mathrsfs}
\usepackage{subcaption}

\newcommand{\Lap}[1]{\mathcal{L} #1}

\newcommand{\beq}{\begin{equation}}
\newcommand{\eeq}{\end{equation}}
\newcommand{\beqa}[1]{\begin{eqnarray}}
\newcommand{\eeqa}{\end{eqnarray}}
\newcommand{\beal}[1]{\begin{align}}
\newcommand{\enal}[1]{\end{align}}
\newcommand{\nn}{\nonumber}

\newcommand{\Tr}{\mbox{Tr}}

\newcommand{\la}{\langle}
\newcommand{\ra}{\rangle}
\newcommand{\commentold}[1]{}

\begin{document}
%
\title{Exact density matrix elements for a driven dissipative system described by a quadratic Hamiltonian}
\author{Sh. Saedi}
\affiliation{Department of Physics, University of Kurdistan, P.O.Box 66177-15175, Sanandaj, Iran}
\author{F. Kheirandish*}
\affiliation{Department of Physics, University of Kurdistan, P.O.Box 66177-15175, Sanandaj, Iran}

\date{\today}

\begin{abstract}
For a prototype quadratic Hamiltonian describing a driven, dissipative system, exact matrix elements of the reduced density matrix are obtained from a generating function in terms of the normal characteristic functions. The approach is based on the Heisenberg equations of motion and operator calculus. The special and limiting cases are discussed.
\end{abstract}



\date{\today}
\maketitle
%
\section{Introduction}\label{Introduction}
Experimental methods in the design of devices used in nanoscale physics and quantum technologies have advanced much in recent years and have led to very high accuracy in measuring instruments. These devices are very sensitive to external potentials and types of noise because their operation is in the domain of quantum mechanics. Therefore, understanding the performance and quantum dynamics of these devices is essential to control them and correct unwanted behaviors. A real quantum device is not an isolated system and interacts with its environment or there may be some external classical sources applied to the system.  Our favorite quantum devices belong to a much wider class of quantum systems, nowadays referred to as open quantum systems \cite{F.1}. The subject of open quantum systems (systems that exchange information with their surroundings) covers a vast range of applications in quantum physics and other related subjects. Generally, by an open quantum system, we mean a possibly driven system as the main system interacting with some other systems modeling its environment. In the terminology of open quantum systems, the main system together with its environment under the influence of external classical sources is considered as a closed system. Therefore, the time-evolution of the total system can be described by a total density matrix evolving unitarily.  If we are interested in the dynamics of the main system or any other subsystem in the environment, the other degrees of freedom should be traced out to get the favorite reduced density matrix. The quantum Brownian motion is an example of an open quantum system that is extensively studied in various branches of physics \cite{F.2,F.3,F.4,F.5,F.6,F.7,F.8,F.9}.
Another important feature of nanoscale quantum devices is their thermodynamical properties. Usually, the quantum fluctuations of the physical quantities in nanoscale quantum devices are of the same order of magnitude as their expectation values leading to a reformulation of thermodynamics in the quantum regime referred to as quantum thermodynamics \cite{G.1,G.2,G.3,G.4,G.5,G.6,G.7,G.8,G.9,G.10}.
There are some other quantum systems whose Hamiltonians resemble the Hamiltonian of the quantum Brownian motion in external sources. The Hamiltonian that we have investigated here is the Hamiltonian given by Eq. (\ref{1}) describing a driven system with a quadratic Hamiltonian $\hat{H}_S$ interacting linearly with its bosonic environment. The Hamiltonian $\hat{H}_S $ appears in many applications in quantum optics \cite{Mandal,Tsai,Piza,H1,H2,H3,H4,H5,H6}.

The quadratic Hamiltonian $\hat{H}_S $ in Eq. (\ref{1}) can be rewritten in terms of the position ($\hat{x}$) and momentum ($\hat{p}$) operator operators, also known as quadrature operators in the terminology of quantum optics, as
\[
\hat{H}_S=\frac{\hat{p}^2}{2m^*}+\frac{1}{2}m^*\omega^{*2}\,\hat{x}^2+\phi_I (\hat{x}\hat{p}+\hat{p}\hat{x}).
\]
The renormalized mass ($m^*$) and frequency ($\omega^*$) are defined by
\begin{eqnarray}
m^* &=& \frac{m}{1-\frac{2\phi_R}{\omega_0}},\nn\\
\omega^* &=& \omega_0\,(1-\frac{4 \phi_R^2}{\omega_0^2}),\nn
\end{eqnarray}
where $\phi_R=\mbox{Re}[\phi]$ and $\phi_I=\mbox{Im}[\phi]$. Therefore, the real part of the complex parameter $\varphi$ is responsible for renormalization of mass and frequency and its imaginary part introduces a term proportional to $\hat{x}\hat{p}+\hat{p}\hat{x}$ to the harmonic oscillator hamiltonian. Here we have implicitly assumed $\phi_R, \phi_I \ll \omega_0$, meaning that the two-boson process is less likely to occur than the one-boson process. From now on we assume that we are considering an oscillator with renormalized mass and frequency so we can set $m=m^*$ and $\omega_0=\omega^*$ and only the imaginary part of $\phi$ is relevant. Adding the terms representing the interaction of the external force $k(t)$ on the oscillator we find the time-dependent Hamiltonian $\hat{H}_{SK} (t)$ as
\begin{eqnarray}
\hat{H}_{SK} (t) &=& \frac{\hat{p}^2}{2m}+\frac{1}{2}m\omega^2\,\hat{x}^2+\phi_I (\hat{x}\hat{p}+\hat{p}\hat{x})+\sqrt{2m\hbar\omega}\,k_R (t)\,\hat{x}+\sqrt{\frac{2\hbar}{m\omega}}\,k_I (t)\,\hat{p},\nn\\
                 &=& \hbar\omega_{0}\,(\hat{a}^\dag \hat{a}+1/2)+\hbar\bar{\phi}\,\hat{a}^2+\hbar\phi\,(\hat{a}^\dag)^2+\hbar\,k(t){\hat{a}^\dag}+\hbar\,\bar{k}(t)\hat{a}.
\end{eqnarray}
The Hamiltonian $\hat{H}_{SK} (t)$ is the Hamiltonian of the system in the absence of a reservoir. The squeezed states generated from the Hamiltonian $\hat{H}_{SK} (t) $ have been investigated in \cite{Basaeia, Toledo}, the Wigner function corresponding to the same hamiltonian is discussed in \cite{Ben}.
The $su(1,1)$ coherent states generated from $\hat{H}_{SK} (t) $ have been studied in \cite{Choi}.
The Hamiltonian $\hat{H}_{SK} (t)$ from the point of view of Fresnel operator has been investigated in \cite{Wang}. Single-mode two-photon systems with Hamiltonian $\hat{H}_{SK} (t)$ have been investigated in \cite{Gilmore}.

There are some approaches to find the reduced density matrix of a subsystem in a combined system, like path integral technique \cite{F.6}, though general, is usually difficult to deal with, Lindblad master equation \cite{F.1} which is based on some approximations, and phenomenological or quantum Langevin approaches \cite{Langevin}. Here, instead, we follow a scheme to find the exact reduced density matrix elements corresponding to the subsystem $\hat{H}_S (t)$ by making intense use of the operator calculus in the Heisenberg picture. Thereby, we indeed find analytic expressions for the generating function of the reduced density matrix. To the best of our knowledge, this approach has not been applied to the Hamiltonian Eq. (\ref{1}) before, and despite its simplicity, could provide closed-form expressions for the reduced density matrix. Knowing the matrix elements of the reduced density matrix, a full description of the dynamics of the main subsystem can be achieved.
\section{The main definitions}\label{sec2}
The prototype system that we have considered in this section is a system described by a quadratic Hamiltonian driven by an external classical source $k(t)\,(\bar{k}(t))$ interacting with a bosonic bath linearly. The total Hamiltonian is
\begin{eqnarray}\label{1}
  \hat{H}=&& \underbrace{\hbar\omega_{0}\,(\hat{a}^\dag \hat{a}+1/2)+\hbar\bar{\phi}\,\hat{a}^2+\hbar\phi\,(\hat{a}^\dag)^2}_{\mbox{Quadratic Hamiltonian}\,\hat{H}_S (t)}+\underbrace{\hbar\,k(t){\hat{a}^\dag}+\hbar\,\bar{k}(t)\hat{a}}_{\mbox{Interaction with external force}\,k(t)}\nn\\
  && +\underbrace{{\sum_{j}\hbar\omega_{j}{\hat{b}^\dag}_{j}\,{\hat{b}}_{j}}}_{\mbox{Bosonic Bath}\,\hat{H}_{R}}
  +\underbrace{\sum_{j}\big[\hbar\,f_{j}{\hat{a}^\dag}{\hat{b}}_{j}+\hbar\,\bar{f}_{j}{\hat{b}^\dag}_{j}{\hat{a}}\big]}_{\mbox{Linear interaction}\,\hat{H}_{SR}},
\end{eqnarray}
where $f_{j}$ are the coupling constants that couple the system to its environment and the parameter $\phi$ is an arbitrary complex parameter. Here, the complex conjugate of any quantity such as $c$ is denoted by $\bar{c}$ and its norm by $|c|$. The Laplace transform of a function is denoted by $\mu(t)$ with $\tilde{\mu}(s)=\Lap[\mu(t)]$ with the inverse $\mu (t)=\mathcal{L}^{-1}[\mu (s)]$.

Our goal is to find the exact matrix components of the reduced density matrix corresponding to the Hamiltonian $\hat{H}_S (t)$. To this end, we first need to find the temporal evolution of the ladder operators. By making use of the Heisenberg equations of motion for the subsystem ladder operators we find (appendix \ref{A})
\begin{equation}\label{2}
  \hat{a}(t)=\alpha_{1}(t)\hat{a}(0)-2\,i\,\phi\,\alpha_{2}(t)\hat{a}^{\dag}(0)-i\sum_{j}M_{j}(t)\hat{b}_{j}(0)\\
  -i\sum_{j}(2\,i\,\phi)\,N_{j}(t)\hat{b}^{\dag}_{j}(0)-i\,\zeta_{1}(t)-i\,(2\,i\,\phi)\zeta_{2}(t),
\end{equation}
\begin{equation}\label{3}
  \hat{a}^{\dag}(t)=\bar{\alpha}_{1}(t)\hat{a}^{\dag}(0)+2\,i\,\bar{\phi}\,\alpha_{2}(t)\hat{a}(0)
  +i\sum_{j}\bar{M}_{j}(t)\hat{b}^{\dag}_{j}(0)-i\sum_{j}(2\,i\,\phi)\bar{N}_{j}(t)\hat{b}_{j}(0)+i\,\bar{\zeta}_{1}(t)-i\,(2\,i\,\phi)\bar{\zeta}_{2}(t),
\end{equation}
where for notational simplicity we have defined the following functions
\begin{align}\label{definitions}
\alpha_{1}(t)=& \mathcal{L}^{-1}\Big[\frac{\tilde{\bar{G}}(s)}{\tilde{L}(s)}\Big],\nn\\
\alpha_{2}(t)=& \mathcal{L}^{-1}\Big[\frac{1}{\tilde{L}(s)}\Big],\nn\\
\tilde{L}(s)=& |\tilde{G}(s)|^{2}-4\left|\phi\right|^{2},\nn\\
\tilde{G}(s)=& s+i\,\omega_{0}+\tilde{\chi}(s),\nn\\
M_{j}(t)=& f_{j}\int_{0}^{t}d\,t'e^{i\,\omega_{j}(t-t')}\alpha_{1}(t'),\nn\\
N_{j}(t)=& \bar{f}_{j}\int_{0}^{t}d\,t'e^{-i\,\omega_{j}(t-t')}\alpha_{2}(t'),\nn\\
\zeta_{1}(t)=& \int_{0}^{t}d\,t'\,\alpha_{1}(t-t')\,k(t'),\nn\\
\zeta_{2}(t)=& \int_{0}^{t}d\,t'\,\alpha_{2}(t-t')\,\bar{k}(t').
\end{align}
Similarly, for the environment ladder operators we find (appendix \ref{A})
\begin{equation}\label{4}
  \hat{b}_{j}(t)=\sum_{k}\left[\Lambda_{jk}(t)\,\hat{b}_{k}(0)+\Lambda'_{jk}(t)\,\hat{b}^{\dag}_{k}(0)-\Gamma_{jk}(t)\,\hat{a}(0)-\Gamma'_{jk}(t)\,\hat{a}^{\dag}(0)-\Omega_{jk}(t)\right],
\end{equation}
\begin{equation}\label{5}
  \hat{b}^{\dag}_{j}(t)=\sum_{k}\left[\bar{\Lambda}_{jk}(t)\,\hat{b}^{\dag}_{k}(0)+\bar{\Lambda'}_{jk}(t)\,\hat{b}_{k}(0)
  -\bar{\Gamma}_{jk}(t)\,\hat{a}^{\dag}(0)-\bar{\Gamma'}_{jk}(t)\,\hat{a}(0)-\bar{\Omega}_{jk}(t)\right],
\end{equation}
where we have defined
\begin{align}\label{definitions2}
\Lambda_{jk}(t)=& e^{-i\,\omega_{j}\,t}\delta_{jk}-\bar{f}_{j}\int_{0}^{t}d\,t'e^{i\,\omega_{j}(t-t')}M_{k}(t'),\nn\\
\Lambda'_{jk}(t)=& -\bar{f}_{j}\int_{0}^{t}d\,t'e^{i\,\omega_{j}(t-t')}(2\,i\,\phi)N_{k}(t'),\nn\\
\Gamma_{jk}(t)=& -i\,\bar{f}_{j}\int_{0}^{t}d\,t'e^{i\,\omega_{j}(t-t')}\alpha_{1k}(t'),\nn\\
\Gamma'_{jk}(t)=& -2\,\phi\bar{f}_{j}\int_{0}^{t}d\,t'e^{i\,\omega_{j}(t-t')}\alpha_{2k}(t'),\nn\\
\Omega_{jk}(t)=& -\bar{f}_{j}\int_{0}^{t}d\,t'e^{i\,\omega_{j}(t-t')}\left(\zeta_{1}(t')-(2\,i\,\phi)\zeta_{2}(t')\right).
\end{align}
In the next section, by making use of the main equations Eqs. (\ref{2}, \ref{3}, \ref{4}, \ref{5}), we will obtain a generating function to produce the reduced density matrix elements of the bosonic mode subsystem.
\section{Reduced density matrix elements}\label{sec3}
According to the terminology of the open quantum systems theory, the whole system described by the Hamiltonian Eq. (\ref{1}) is a closed system having a unitary time-evolution given by
\begin{equation}\label{7}
  \hat{\rho}(t)=\hat{U}(t)\,\hat{\rho}(0)\,\hat{U}^{\dag}(t),
\end{equation}
where the initial density matrix of the whole system ($\hat{\rho} (0)$) is usually assumed to be a separable state
\begin{equation}\label{6}
  \hat{\rho}(0)=\hat{\rho}_{S}(0)\otimes\hat{\rho}_{R}(0).
\end{equation}
The reduced density matrix of the bosonic-mode subsystem can be obtained by tracing out the degrees of freedom of the environment
\begin{equation}\label{8}
  \hat{\rho}_{S}(t)=\Tr_{R}\left\{\hat{\rho}(t)\right\}.
\end{equation}
We are interested in the matrix elements of the reduced density matrix. We have
\begin{align}\label{9}
   \langle\,n\,|\hat{\rho}_{S}(t)|\,m\,\rangle &=\langle\,n\,|Tr_{R}\left\{\hat{U}(t)\hat{\rho}(t_{0})\hat{U}^{\dag}(t)\right\}|\,m\,\rangle,\nn\\
   &=\left(|\,m\,\rangle\langle\,n\,|Tr_{R}\left\{\hat{U}(t)\hat{\rho}(t_{0})\hat{U}^{\dag}(t)\right\}\right),\nn\\
   &=Tr\left\{\left(|\,m\,\rangle\langle\,n\,|\otimes\,\hat{I}_{R}\right)\hat{U}(t)\hat{\rho}(t_{0})\hat{U}^{\dag}(t)\right\},\nn\\
   &=Tr\left\{\overbrace{\hat{U}^{\dag}(t)\left(|\,m\,\rangle\langle\,n\,|\otimes\,\hat{I}_{R}\right)\hat{U}(t)}^{\hat{Q}_{nm}}\hat{\rho}(t_{0})\right\},
\end{align}
therefore,
\begin{align}\label{10}
   \langle\,n\,|\hat{\rho}_{S}(t)|\,m\,\rangle&=Tr\left\{\hat{Q}_{nm}\,\hat{\rho}(0)\right\}\nn \\
   &=\Tr\left\{\hat{Q}_{nm}\,\hat{\rho}_{S}(0)\otimes\hat{\rho}_{R}(0)\right\}.
\end{align}
In Eq. (\ref{9}), the operator $\hat{I}_{R}$ is the identity operator over the environment Hilbert space. The matrix elements $\hat{Q}_{mn}$ can be written in terms of the ladder operators in the Heisenberg representation (appendix \ref{B})
\begin{equation}\label{11}
  \hat{Q}_{nm}=\frac{1}{\sqrt{m!\,n!}}\sum_{s=0}^{\infty}\frac{(-1)^{s}}{s!}\left(\hat{a}^{\dag}(t)\right)^{m+s}\left(\hat{a}(t)\right)^{n+s},
\end{equation}
where
\begin{align}\label{definitions3}
\hat{a}(t)=& \hat{C}(t)-i\left(\hat{B}(t)+\zeta(t)\right),\nn\\
\hat{C}(t)=& \alpha_{1}(t)\hat{a}(0)-2\,i\,\phi\,\alpha_{2}(t)\hat{a}^{\dag}(0),\nn\\
\hat{B}(t)=& \sum_{j}\left(M_{j}(t)\hat{b}_{j}(0)+2\,i\,\phi\,N_{j}(t)\hat{b}^{\dag}_{j}(0)\right),\nn\\
\zeta(t)=& \left(\zeta_{1}(t)+2\,i\,\phi\,\zeta_{2}(t)\right),
\end{align}
and $\hat{a}^\dag (t)$ can be obtained by taking the hermitian conjugation of the relations defined in Eq. (\ref{definitions3}). By inserting the expressions for $\hat{a}(t)$ and $\hat{a}^\dag (t)$ into Eq. (\ref{11}), one easily finds
\begin{equation}\label{12}
  \hat{Q}_{nm}=\frac{1}{\sqrt{m!\,n!}}\sum_{s=0}^{\infty}\frac{(-1)^{s}}{s!}
  \Tr\left\{\left(\hat{C}^{\dag}(t)+i\left(\hat{B}^{\dag}(t)+\bar{\zeta}(t)\right)\right)^{m+s}
  \left(\hat{C}(t)-i\left(\hat{B}(t)+\zeta(t)\right)\right)^{n+s}\right\}.
\end{equation}
Now by making use of Eq. (\ref{10}) we have
\begin{align}\label{13}
  \langle\,n\,|\hat{\rho}_{S}(t)|\,m\,\rangle=\frac{(-1)^n}{\sqrt{m!\,n!}}\sum_{s=0}^{\infty}\frac{1}{s!}\frac{\partial^{m+s}}
 {\partial\lambda^{m+s}}\frac{\partial^{n+s}}{\partial(\bar{\lambda})^{n+s}}
   \Tr\left\{e^{\lambda\left(\hat{C}^{\dag}(t)+i\left(\hat{B}^{\dag}(t)+\bar{\zeta}(t)\right)\right)}
 e^{-\bar{\lambda}\left(\hat{C}(t)-i\left(\hat{B}(t)+\zeta(t)\right)\right)}\hat{\rho}_{S}(0)\otimes\hat{\rho}_{R}(0)\right\}_{\lambda=\bar{\lambda}=0}.
\end{align}
From the definitions of operators $\hat{B}$ and $\hat{C}$, we observe that
\begin{equation}\label{BC}
[\hat{B},\hat{C}]=[\hat{B},\hat{C}^{\dag}]=0,
\end{equation}
so we can rewrite Eq. (\ref{13}) as
\begin{align}\label{14}
 & \langle\,n\,|\hat{\rho}_{S}(t)|\,m\,\rangle =\nn\\
 & \frac{(-1)^n}{\sqrt{m!\,n!}}\sum_{s=0}^{\infty}\frac{1}{s!}
 \frac{\partial^{m+s}}{\partial\lambda^{m+s}}\frac{\partial^{n+s}}{\partial(\bar{\lambda})^{n+s}}\,
   \Bigg[e^{i\,\lambda\,\bar{\zeta}(t)+i\,\bar{\lambda}\zeta(t)}
 \overbrace{\Tr_{S}\left\{e^{\lambda\hat{C}^{\dag}(t)}e^{-\bar{\lambda}\hat{C}(t)}\hat{\rho}_{S}(0)\right\}}^{I_{\hat{C}}}
 \overbrace{\Tr_{R}\left\{e^{i\,\lambda\hat{B}^{\dag}(t)}e^{i\,\bar{\lambda}\hat{B}(t)}\hat{\rho}_{R}(0)\right\}}^{I_{\hat{B}}}\Bigg]_{\lambda=\bar{\lambda}=0},\nn\\
 & =\frac{(-1)^n}{\sqrt{m!\,n!}}\frac{\partial^{m}}
  {\partial\lambda^{m}}\frac{\partial^{n}}{\partial(\bar{\lambda})^{n}}\,e^{\partial_{\lambda}\partial_{\bar{\lambda}}}\Big[e^{i\,\lambda\,\bar{\zeta}(t)+i\,\bar{\lambda}\zeta(t)}
  I_{\hat{C}}\,I_{\hat{B}}\Big]_{\lambda=\bar{\lambda}=0}.
\end{align}
Eq. (\ref{14}) is a general result giving the components of the reduced density matrix in terms of a generating function. Note that $I_{\hat{C}}$ and $I_{\hat{B}}$ are normal characteristic functions in the terminology of quantum optics. To proceed, let us assume that the initial state of the environment is a thermal state with temperature $T$
\begin{align}\label{15}
   \hat{\rho}_{R}(t) &=\frac{1}{Z_{R}}\prod_{j}\,e^{-\beta\hbar\omega_{j}\hat{b}^{\dag}_{j}\hat{b}_{j}},\nn\\
  Z_{R} &=\prod_{j}z_{j},\nn\\
  z_{j} &=\Tr_{j}\left\{e^{-\beta\hbar\omega_{j}\hat{b}^{\dag}_{j}\hat{b}_j}\right\},
\end{align}
where $\beta=1/\kappa_B T$ and $\kappa_B$ is the Boltzmann constant. Also, $\Tr_{j}$ denotes the trace over the base vectors corresponding to the $j$th oscillator of the environment. One can obtain $I_{\hat{B}}$ easily as (appendix \ref{C})
\begin{equation}\label{16}
 \Tr_{R}\left\{e^{i\,\lambda\,\hat{B}^{\dag}(t)}e^{i\,\bar{\lambda}\,\hat{B}(t)}{\hat{\rho}}_{R}(0)\right\}=e^{\vartheta[\lambda,\bar{\lambda},t]},
\end{equation}
where
\begin{align}\label{17}
\vartheta[\lambda,\bar{\lambda},t]=& \sum\limits_{k}\Big[-\lambda\,\bar{\lambda}\Big(\frac{|V_{k}(t)|^{2}}{e^{\beta\hbar\,\omega_{k}}-1}+4|\phi|^2\,|N_k (t)|^2\Big)+i(\lambda^2\,\bar{\phi}\bar{N}_k(t) \bar{M}_k(t)-\bar{\lambda}^2 \phi N_k(t) M_k(t))\Big],\\
  V_{k}(t)=& M_{k}(t)+2\,i\,\bar{\phi}\,\bar{N}_{k}(t),
\end{align}
therefore,
\begin{align}\label{18}
  \langle\,n\,|\hat{\rho}_{S}(t)|\,m\,\rangle=&\frac{(-1)^n}{\sqrt{m!\,n!}}\sum_{s=0}^{\infty}\frac{1}{s!}\frac{\partial^{m+s}}
  {\partial\lambda^{m+s}}\frac{\partial^{n+s}}{\partial(\bar{\lambda})^{n+s}}
  \bigg[e^{i\,\lambda\,\bar{\zeta}(t)+i\,\bar{\lambda}\zeta(t)}e^{\vartheta[\lambda,\bar{\lambda},t]}
  \,\Tr_{S}\left\{e^{\lambda\hat{C}^{\dag}(t)}e^{-\bar{\lambda}\hat{C}(t)}\hat{\rho}_{S}(0)\right\}\bigg]_{\lambda=\bar{\lambda}=0},\nn\\
  =& \frac{(-1)^n}{\sqrt{m!\,n!}}\frac{\partial^{m}}
  {\partial\lambda^{m}}\frac{\partial^{n}}{\partial(\bar{\lambda})^{n}}\,e^{\partial_{\lambda}\partial_{\bar{\lambda}}}
  \bigg[e^{i\,\lambda\,\bar{\zeta}(t)+i\,\bar{\lambda}\zeta(t)}e^{\vartheta[\lambda,\bar{\lambda},t]}
  \,\Tr_{S}\left\{e^{\lambda\hat{C}^{\dag}(t)}e^{-\bar{\lambda}\hat{C}(t)}\hat{\rho}_{S}(0)\right\}\bigg]_{\lambda=\bar{\lambda}=0}.
\end{align}
Eq. (\ref{18}) is our main result, giving the reduced density matrix elements using a generating function. In the next section, as an application of the main result, we assume that the bosonic mode is initially prepared in a coherent state.
\section{The system is initially prepared in a coherent state}\label{sec4}
%
As an application of the general formula Eq. (\ref{18}), let us assume that the initial state of the main system is a coherent state
\begin{equation}\label{19}
  \hat{\rho}_{S}(0)=|\gamma\rangle\langle\gamma|,
\end{equation}
in this case, the normal characteristic function $I_{\hat{C}}$ can be obtained as (appendix \ref{D})
\begin{align}\label{20}
 \Tr_{S}\left\{e^{\lambda\hat{C}^{\dag}(t)}e^{-\bar{\lambda}\hat{C}(t)}\hat{\rho}_{S}(0)\right\}= e^{\alpha_{2}(t)\left(i\,\lambda^{2}\bar{\phi}\bar{\alpha}_{1}(t)-i\,\bar{\lambda}^{2}\phi\alpha_{1}(t)
 -4\lambda\bar{\lambda}|\phi|^{2}\alpha_{2}(t)\right)}
  e^{\bar{\gamma}\left(\lambda\bar{\alpha}_{1}(t)+2\,i\,\bar{\lambda}\phi\alpha_{2}(t)\right)}
  e^{-\gamma\left(\bar{\lambda}\alpha_{1}(t)-2\,i\,\lambda\bar{\phi}\alpha_{2}(t)\right)},
\end{align}
and Eq. (\ref{18}) can be rewritten as
\begin{align}\label{21}
& \langle\,n\,|\hat{\rho}_{S}(t)|\,m\,\rangle=\frac{(-1)^n}{\sqrt{m!\,n!}}\frac{\partial^{m}}
  {\partial\lambda^{m}}\frac{\partial^{n}}{\partial(\bar{\lambda})^{n}}\,e^{\partial_{\lambda}\partial_{\bar{\lambda}}}\nn\\
  &\left[e^{i\left(\lambda\bar{\zeta}(t)+\bar{\lambda}\zeta(t)\right)}
    e^{-4\lambda\bar{\lambda}|\phi|^{2}\alpha_{2}^{2}(t)+i\alpha_{2}(t)\left(\lambda^{2}\bar{\phi}\bar{\alpha}_{1}(t)-\bar{\lambda}^{2}\phi\alpha_{1}(t)\right)}
  e^{\vartheta[\lambda,\bar{\lambda},t]}
  e^{\left(\lambda\bar{\gamma}\bar{\alpha}_{1}(t)-\bar{\lambda}\gamma\alpha_{1}(t)\right)
  +2\,i\,\alpha_{2}(t)\left(\bar{\lambda}\bar{\gamma}\phi+\lambda\,\gamma\,\bar{\phi}\right)}\right]_{\lambda=\bar{\lambda}=0}.
\end{align}
Therefore, if the initial state of the main system is a coherent state and the initial state of the environment is a Maxwell Boltzmann thermal state then the elements of the reduced density matrix can be obtained from a generating function given by Eq. (\ref{21}). If we set ($\phi=0$), the diagonal elements of the reduced density matrix $P_n (t)=\langle\,n\,|\hat{\rho}_{S}(t)|\,n\,\rangle$ are
\begin{equation}\label{22}
  P_{n}(t)|_{\phi=0}=\frac{(-1)^n}{n!}\Big(\frac{\partial}{\partial\lambda}\frac{\partial}{\partial\bar{\lambda}}\Big)^{n}
  \overbrace{\sum_{s=0}^{\infty}\frac{1}{s!}\Big(\frac{\partial}{\partial\lambda}\frac{\partial}{\partial\bar{\lambda}}\Big)^{s}}
  ^{e^{\partial_{\lambda}\partial_{\bar{\lambda}}}}I\Bigg|_{\lambda=\bar{\lambda}=0},
\end{equation}
where
\begin{align}\label{23}
   I=& e^{-\lambda\bar{\lambda}\eta(t)+\lambda\,\bar{Z}-\bar{\lambda}\,Z},\nn\\
  \eta(t) =& \sum\limits_{k}\frac{|M_{k}(t)|^{2}}{e^{\beta\hbar\,\omega_{k}}-1},\nn\\
   Z=& -i\zeta(t)+\gamma\alpha_{1}(t),\nn\\
   \bar{Z}=& i\bar{\zeta}(t)+\bar{\gamma}\bar{\alpha}_{1}(t).
\end{align}
Therefore, (appendix \ref{E})
\begin{equation}\label{24}
 P_{n}(t)|_{\phi=0}=\frac{e^{-\frac{\left|Z\right|^{2}}{1+\eta(t)}}}{1+\eta(t)}\left(\frac{\eta(t)}{\eta(t)+1}\right)^{n}
 L_{n}\left(\frac{-|Z|^{2}}{\eta(t)(1+\eta(t))}\right),
\end{equation}
where $L_n [x]$ is a Laguerre polynomial of degree $n$. From Eq. (\ref{24}) the mean excitation number $\bar{n}$ at temperature $T$ and time $t$ is
\begin{equation}\label{meann}
\bar{n}_{T}(t)=\la n\ra_{T}(t)=\sum_{n=0}^\infty n\,P_n (t)=|Z(t)|^2+\eta(t),
\end{equation}
at zero temperature we have $\eta(t)\rightarrow 0$, so $|Z(t)|^2=\bar{n}_{0}(t)$, therefore, $\bar{n}_{T}(t)-\bar{n}_{0}(t)=\eta (t)$.
If we set ($\phi=0$) then in low temperature regime ($T\rightarrow 0$), we have
\begin{align}\label{noKerr}
& \eta(t)\rightarrow 0,\nn\\
& V_{k}(t)=M_{k}(t),\nn\\
& \tilde{L}(s)=\left|\tilde{G}(s)\right|^{2}\Longrightarrow\alpha_{1}(t)=\mathcal{L}^{-1}\left[\frac{1}{\tilde{G}(s)}\right],\nn\\
& \tilde{G}(s)=s+i\,\omega_{0}+\tilde{\chi}(s).
\end{align}
and by making use of the identity
\beq\label{Math}
\lim_{\eta\rightarrow 0}\,\left(\frac{\eta(t)}{\eta(t)+1}\right)^{n}
 L_{n}\left(\frac{-|Z|^{2}}{\eta(t)(\eta(t)+1)}\right)=\frac{|Z|^{2n}}{n!},
\eeq
we deduce
\begin{equation}\label{28}
 P_{n}(t)|_{\phi=0}=\frac{e^{-|Z|^{2}}|Z|^{2n}}{n!}=
 \frac{|\gamma\alpha_{1}(t)-i\zeta_{1}(t)|^{2n}e^{-|\gamma\alpha_{1}(t)-i\zeta_{1}(t)|^{2}}}{n!},
\end{equation}
which is a Poisson distribution with the mean number parameter $\langle\,n\,\rangle$ given by
\begin{equation}\label{29}
  \left\langle\,n\,\right\rangle=|\gamma\alpha_{1}(t)-i\zeta_{1}(t)|^{2}.
\end{equation}
%
\subsection{Example}\label{Example1}
%
For the choice $\phi_I=0$, $k(t)=k_0\sin(\nu t)$ and the memory-less response function $\chi(t)=\chi_0\,\delta(t)$, we find in the large-time limit ($t\gg \chi_0^{-1}$) the following time-independent values
\begin{eqnarray}
&& \alpha_1 (t) \mapsto 0,\nn\\
&& |Z(t)|^2_{\phi=0}=|\zeta_1 (t)|^2 \mapsto \frac{k_0^2 [8\nu^2+2 (\chi_0^2+4 \omega_0^2)]}{(4\nu^2+\chi_0^2-4\omega_0^2)^2+16\chi_0^2\,\omega_0^2},\nn\\
&& \eta(t) \mapsto \sum_{k} \frac{4 |f_k|^2}{\chi_0^2+4 (\omega_0+\omega_k)^2}\,\frac{1}{e^{\frac{\hbar\omega_k}{k_B T}-1}}.\nn
\end{eqnarray}
Note that $\eta$ is a temperature-dependent parameter. The probability $P_n (\eta)$ for $n=0,1,2,3$ is depicted in Fig.1 in terms of the dimensionless parameter $\eta$. The most probable excitation value (in zero temperature) belongs to $n=\bar{n}_0$ which for the values assigned to the parameters $\omega_0$, $\chi_0$, $k_0$ and $\nu$ in the caption of the Fig.1 is $n=0$. If we increase the strength of the external source for example by choosing the values $k_0=0.2\,\omega_0$, $\chi_0=0.1\,\omega_0$, and $\nu=0.99\,\omega_0$, then we will find the results as depicted in Fig.2 for the values $n=0, 1, 2, 3, 4$. It is seen that the most probable value (at zero temperature) corresponds to $n=[\bar{n}_0]=3$ where $[a]$ returns the integer part of $a$. Note that, in large-time limit and finite temperature we have $\bar{n}_T-\bar{n}_0=\eta(T)$. The results are consistent with our expectations and the results known in the literature.
\begin{figure}
  \centering
  \includegraphics[width=9cm]{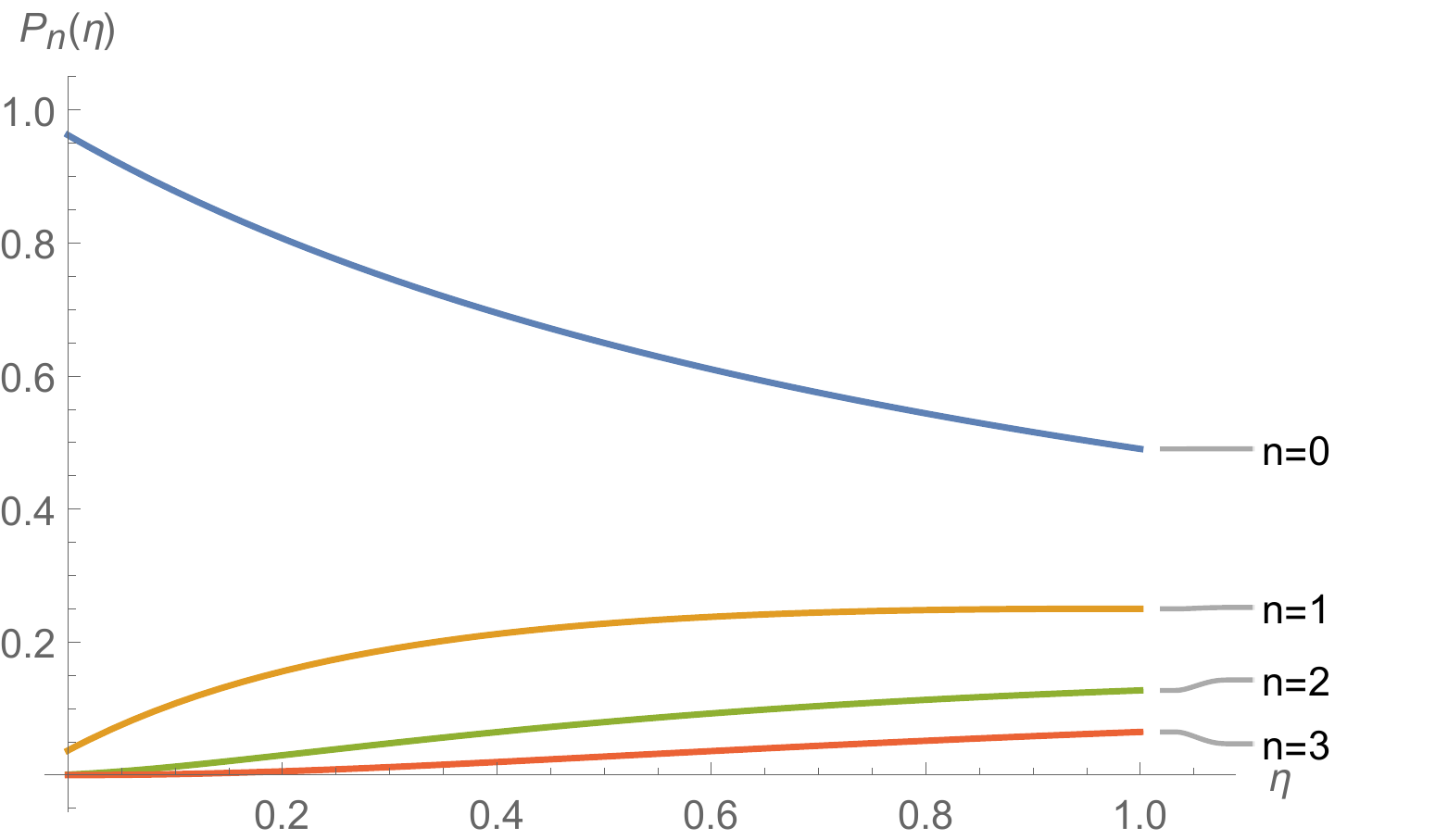}
  \caption{(Color online) The probability $P_n (\eta)$ (see Eq.(\ref{24})) for $n=0,1,2,3$ is depicted for the values $k_0=0.02\,\omega_0$, $\chi_0=0.1\,\omega_0$, and $\nu=0.99\,\omega_0$ in terms of the dimensionless variable $\eta$ in large-time limit.}\label{Fig1}
\end{figure}
\begin{figure}
  \centering
  \includegraphics[width=9cm]{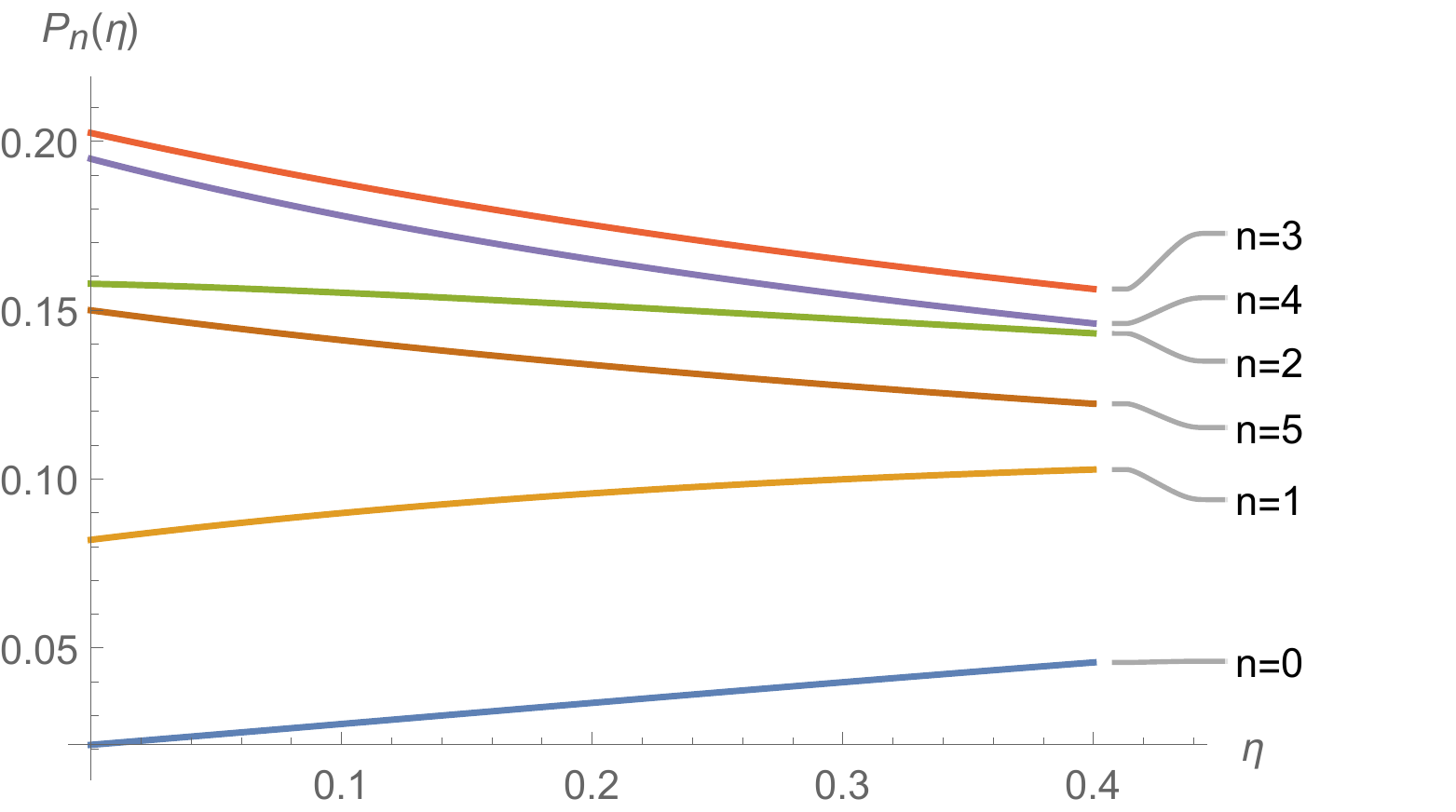}
  \caption{(Color online) The probability $P_n (\eta)$ (see Eq.(\ref{24})) for $n=0,1,2,3$ is depicted for the values $k_0=0.2\,\omega_0$, $\chi_0=0.1\,\omega_0$, and $\nu=0.99\,\omega_0$ in terms of the dimensionless variable $\eta$ in large-time limit.}\label{Fig1}
\end{figure}
%
\section{Strong coupling with external source and low dissipation regime}
Let us assume that the system is initially prepared in the ground state $\rho_S (0)=|0\ra\la 0|$, then in the absence of dissipation, by setting $n=m$ and $\gamma=0$  in Eq. (\ref{21}) we find
\begin{equation}\label{L1}
P_n (t)=\frac{(-1)^n}{n!}(\partial_\lambda \partial_{\bar{\lambda}})^n\,e^{\partial_{\lambda}\partial_{\bar{\lambda}}} \left[e^{i\left(\lambda\bar{\zeta}(t)+\bar{\lambda}\zeta(t)\right)}
    e^{-4\lambda\bar{\lambda}\phi_I^{2}\alpha_{2}^{2}(t)+\alpha_{2}(t)\phi_I\left(\lambda^{2}\bar{\alpha}_{1}(t)+\bar{\lambda}^{2}\alpha_{1}(t)\right)}
  \right]_{\lambda=\bar{\lambda}=0}.
\end{equation}
To simplify the calculations, we ignore from the term proportional to $\phi_I^2$ in the exponential term in Eq. (\ref{L1}) ($\phi_I\ll\omega_0$), in this case the exponential term is separable in terms of $\lambda,\,\bar{\lambda}$, therefore, by expanding $\exp(\partial_{\lambda}\partial_{\bar{\lambda}})$ we have
\begin{eqnarray}
P_n (t) &=& \frac{(-1)^n}{n!}\sum_{s=0}^\infty \frac{1}{s!}\Big(\partial_\lambda^{n+s}e^{\big(i\lambda\bar{\zeta}(t)+\alpha_2 (t)\phi_I\bar{\alpha}_1\lambda^2\big)}\Big)\Big(\partial_{\bar{\lambda}}^{n+s}e^{\big(i\bar{\lambda}\zeta(t)+\alpha_2 (t)\phi_I\alpha_1\bar{\lambda}^2\big)}\Big)\Big|_{\lambda=\bar{\lambda}=0}.
\end{eqnarray}
Now using the generating function of Hermite polynomials
\begin{equation}\label{L2}
e^{-t^2+2tx}=\sum_{s=0}^\infty \frac{t^k}{k!}\,H_k (x),
\end{equation}
and changing the variable $\lambda=y/\sqrt{i\alpha_2 (t)\phi_I \bar{\alpha}_1}$, one easily finds
\begin{equation}\label{L3}
P_n (t)=\frac{(\alpha_2 (t)\phi_I |\alpha_1 (t)|)^n}{n!}\sum_{s=0}^\infty \frac{(-\alpha_2 (t)\phi_I |\alpha_1 (t)|)^s}{s!}\Big |H_{n+s} \Big(\frac{-\zeta(t)}{2\sqrt{\alpha_2 (t)\phi_I\alpha_1}}\Big)\Big |^2.
\end{equation}
By making use of Eqs. (\ref{definitions},\ref{definitions3}) we have
\begin{eqnarray}\label{L4}
&&\tilde{ \bar{G}}(s)=s-i\omega_0,\nn\\
&& \tilde{L}(s)=s^2+\omega_0^2-4\phi_I^2\approx s^2+\omega_0^2,\nn\\
&& \alpha_1 (t)=e^{-i\omega_0 t}\rightarrow|\alpha_1 (t)|=1,\nn\\
&& \alpha_2 (t)= \frac{\sin(\omega_0 t)}{\omega_0},\nn\\
&& \zeta(t)=\zeta_1 (t)-2\phi_I\zeta_2 (t).
\end{eqnarray}
Note that at the times $\tau=m\,\pi\,\,(m=1,2,3,\cdots) $ we have $\alpha_2 (t)=0$ and equation Eq. (\ref{L3}) becomes singular at these points, but, these singular points are removable and one can easily show that at this times Eq. (\ref{L3}) tends to a poisson distribution given by
\begin{equation}\label{L5}
P_n (t=m\,\pi)=\frac{(\lambda_m)^n}{n!}\,e^{-\lambda_m^2},\,\,m=1,2,3,\cdots,
\end{equation}
where
\begin{equation}
\lambda_m=|\zeta(t=m\pi)|^2.
\end{equation}
For the external source $k(t)=k_0\,\sin(\nu t)$, from Eqs. (\ref{definitions},\ref{L4}) we have
\begin{eqnarray}\label{example1}
&& \zeta_1 (t)=\frac{k_0[\nu\,e^{-i\omega_0 t}-\nu\,\cos(\nu t)+i\omega_0\,\sin(\nu t)]}{\nu^2-\omega_0^2},\nn\\
&& \zeta_2 (t)=\frac{k_0[\omega_0\,\sin(\nu t)-\nu\,\sin(\omega_0 t)]}{\omega_0 (\omega_0^2-\nu^2)},\nn\\
&& \zeta(t)=\frac{k_0[\nu\omega_0 e^{-i\omega_0 t}-\nu\omega_0\cos(\nu t)+(2\phi_I/\omega_0+i)\omega_0^2\sin(\nu t)-2\nu\phi_I\,\sin(\omega_0 t)]}{\omega_0(\nu^2-\omega_0^2)}.
\end{eqnarray}
The probability $P_n (\tau)$ for $n=0,1,2,3,4$ is depicted for the values $k_0=\omega_0$, $\phi_I=0.1\,\omega_0$, and $\nu=0.9\,\omega_0$ in terms of the dimensionless variable $\tau=\omega_0 t$ in Fig. 3. Note the order of excitations in time ($n=1,\, n=2,\, n=3,\, n=4$) as we expected.
\begin{figure}
  \centering
  \includegraphics[width=9cm]{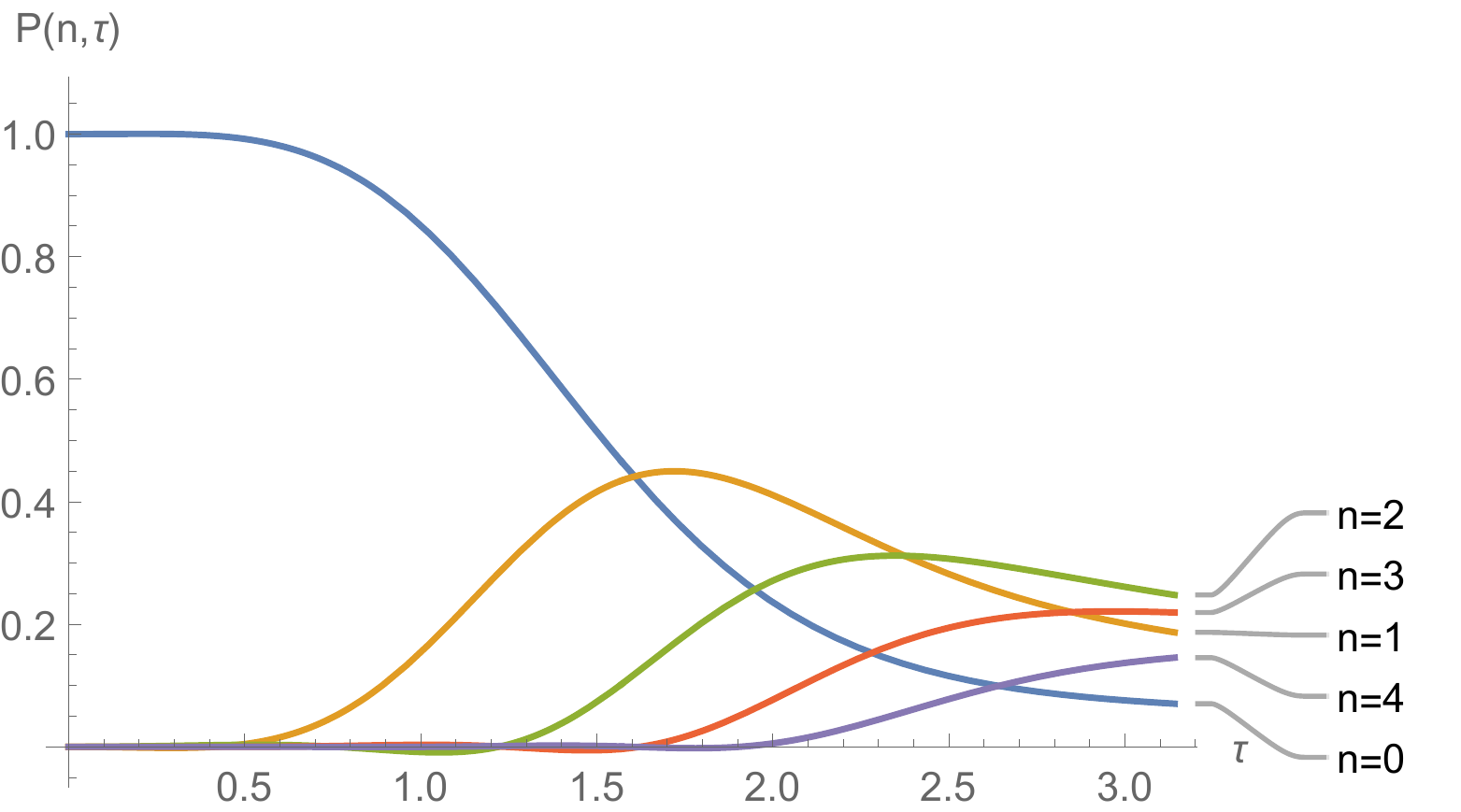}
  \caption{(Color online) The probability $P_n (\tau)$ (see Eq.(\ref{L3})) for $n=0,1,2,3,4$ is depicted for the values $k_0=\omega_0$, $\phi_I=0.1\,\omega_0$, and $\nu=0.9\,\omega_0$ in terms of the dimensionless variable $\tau=\omega_0 t$}\label{Fig1}
\end{figure}
%
\section{Conclusion}\label{sec6}
We have considered a driven, dissipative quantum system described by a time-dependent quadratic Hamiltonian and found a generating function Eq. (\ref{14}) to find the exact matrix elements of the reduced density matrix. The generating function is given in terms of the well-known normal characteristic functions in the terminology of quantum Optics. For the case of a thermal environment with a Maxwell-Boltzmann equilibrium state, an exact expression for the components of the reduced density matrix is obtained given by Eq. (\ref{18}). Explicit expressions for reduced density matrix components are obtained when the subsystem is initially prepared in a coherent state. Despite the simplicity of the method, while deriving the main result Eq. (\ref{18}), assumptions like weak or strong coupling and/or Markovian approximation have not been applied.
%
	

\appendix	
\section{}\label{A}
By making use of the Heisenberg equations of motion we have
\begin{align}\label{A.1}
  \hat{\dot{a}}=\frac{1}{i\,\hbar}\left[\hat{a},\hat{H}\right]=-i\,\omega_{0}\hat{a}-i\sum_{j}f_{j}\hat{b}_{j}-i\,k(t)-2\,i\,\phi\hat{a}^{\dag},
\end{align}
\begin{align}\label{A.3}
  \hat{\dot{b}}_{j}=\frac{1}{i\,\hbar}\left[\hat{b}_{j},\hat{H}\right]=-i\,\omega_{j}\hat{b}_{j}-i\,\bar{f}_{j}\hat{a}.
\end{align}
Eq. (\ref{A.3}) can be solved formally as
\begin{equation}\label{A.9}
 \hat{b}_{j}(t)=e^{-i\omega_{j}t}\,\hat{b}_{j}(0)\,-i\,\bar{f}_{j}\int_{0}^{t}\,dt'\,e^{-i\omega_{j}(t-t')}\,\hat{a}(t'),
\end{equation}
by inserting Eq. (\ref{A.9}) into Eq. (\ref{A.1}) we will find
\begin{equation}\label{A.13}
 \hat{\dot{a}}+i\,\omega_{0}\hat{a}+\int_{0}^{t}dt'\,\chi(t-t')\,\hat{a}(t')
 =-i\sum_{j}f_{j}e^{-i\omega_{j}t}\hat{b}_{j}(0)-i\,k(t)-2\,i\,\phi\hat{a}^{\dag}(t),
\end{equation}
where
\begin{equation}\label{Ki}
 \chi(t-t')=\sum_{j}|{f}_{j}|^{2}e^{-i\omega_{j}(t-t')},
\end{equation}
is the response function or the memory function of the medium. By taking the Laplace transform of both sides of Eq. (\ref{A.13}) we have
\begin{align}\label{A.15}
 \tilde{a}(s)=&\left[\frac{1}{s+i\,\omega_{0}+\tilde{\chi}(s)}\right]\hat{a}(0)
  -\left[\frac{i}{s+i\,\omega_{0}+\tilde{\chi}(s)}\right]\sum_{j}{f}_{j}\left[\frac{1}{s+i\,\omega_{j}}\right]\hat{b}_{j}(0)\nn \\
 &-\left[\frac{2\,i\,\phi}{s+i\,\omega_{0}+\tilde{\chi}(s)}\right]\tilde{a}^{\dag}(s)-\left[\frac{i}{s+i\,\omega_{0}+\tilde{\chi}(s)}\right]\tilde{k}(s),
\end{align}
where
\begin{equation}\label{Kis}
  \tilde{\chi}(s)=\mathcal{L}\Big[\sum_{j}|{f}_{j}|^{2}e^{-i\omega_{j}t}\Big].
\end{equation}
Now we can rewrite Eq. (\ref{A.15}) and its adjoint as
\begin{equation}\label{A.17}
  \tilde{a}(s)=\left[\frac{1}{\bar{G}(s)}\right]\hat{a}(0)-\left[\frac{i}{\tilde{G}(s)}\right]\sum_{j}\left[\frac{{f}_{j}}{s+i\,\omega_{j}}\right]\hat{b}_{j}(0)
  -\left[\frac{2\,i\,\phi}{\tilde{G}(s)}\right]\tilde{a}^{\dag}(s)-\left[\frac{i}{\tilde{G}(s)}\right]\tilde{k}(s),
\end{equation}
\begin{equation}\label{A.18}
  \tilde{a}^{\dag}(s)=\left[\frac{1}{\tilde{\bar{G}}(s)}\right]\hat{a}^{\dag}(0)+\left[\frac{i}{\tilde{\bar{G}}(s)}\right]
  \sum_{j}\left[\frac{{\bar{f}}_{j}}{s-i\,\omega_{j}}\right]\hat{b}^{\dag}_{j}(0)
 +\left[\frac{2\,i\,\bar{\phi}}{\tilde{\bar{G}}(s)}\right]\tilde{a}(s)+\left[\frac{i}{\tilde{\bar{G}}(s)}\right]\tilde{\bar{k}}(s),
\end{equation}
leading to
\begin{align}\label{A.19}
   \tilde{a}(s)=&\left[\frac{\tilde{\bar{G}}(s)}{\tilde{L}(s)}\right]\hat{a}(0)-\left[\frac{2\,i\,\phi}{\tilde{L}(s)}\right]\hat{a}^{\dag}(0)
  -\left[\frac{i\tilde{\bar{G}}(s)}{\tilde{L}(s)}\right]\sum_{j}\left[\frac{{f}_{j}}{s+i\,\omega_{j}}\right]\hat{b}_{j}(0)\nn \\
  &+\left[\frac{2\,\phi}{\tilde{L}(s)}\right]\sum_{j}\left[\frac{{\bar{f}}_{j}}{s-i\,\omega_{j}}\right]\hat{b}^{\dag}_{j}(0)
  -\left[\frac{i\tilde{\bar{G}}(s)}{\tilde{L}(s)}\right]\tilde{k}(s)+\left[\frac{2\,\phi}{\tilde{L}(s)}\right]\tilde{\bar{k}}(s).
\end{align}
Now using the inverse Laplace transform we find
\begin{equation}\label{A.22}
 \hat{a}(t)=\alpha_{1}(t)\hat{a}(0)-2\,i\,\phi\alpha_{2}(t)\hat{a}^{\dag}(0)-i\sum_{j}M_{j}(t)\hat{b}_{j}(0)
  -i\sum_{j}\left(2\,i\,\phi\right)N_{j}(t)\hat{b}^{\dag}_{j}(0)-i\,\zeta_{1}(t)-i\left(2\,i\,\phi\right)\,\zeta_{2}(t),
\end{equation}
and from Eq. (\ref{A.9}) we deduce
\begin{equation}\label{A.24}
 \hat{b}_{j}(t)=\sum_{k}\left[\Lambda_{jk}(t)\hat{b}_{k}(0)+\Lambda'_{jk}(t)\hat{b}^{\dag}_{k}(0)
 -\Gamma_{jk}(t)\hat{a}(0)-\Gamma'_{jk}(t)\hat{a}^{\dag}(0)-\Omega_{jk}(t)\right].
\end{equation}
%
\section{}\label{B}
We have
\begin{align}\label{B.1}
 \hat{Q}_{nm}&=\hat{U}^{\dag}(t)\left(|m\,\rangle\,\langle\,n|\otimes\,I_{R}\right)\hat{U}(t),\nn\\
   &=\hat{U}^{\dag}(t)\frac{\hat{a}^{\dag}(0)^{m}}{\sqrt{m!}}|0\,\rangle\langle\,0|\frac{\hat{a}(0)^{n}}{\sqrt{n!}}\otimes\,I_{R}\hat{U}(t),
\end{align}
and \cite{Louisell}
\begin{equation}\label{B.2}
    |0\,\rangle\langle\,0|=\sum_{s=0}^{\infty}\frac{(-1)^{s}}{s!}\left(\hat{a}^{\dag}(0)\right)^{s}\left(\hat{a}(0)\right)^{s},
\end{equation}
by inserting Eq. (\ref{B.2}) into Eq. (\ref{B.1}) we obtain
\begin{align}\label{B.3}
 \hat{Q}_{nm}=&\frac{1}{\sqrt{m!n!}}\sum_{s=0}^{\infty}\frac{(-1)^{s}}{s!}\hat{U}^{\dag}(t)\left(\hat{a}^{\dag}(0)\right)^{m+s}
 \left(\hat{a}(0)\right)^{n+s}\otimes\,I_{R}\,\hat{U}(t),\nn\\
 =& \frac{1}{\sqrt{m!n!}}\sum_{s=0}^{\infty}\frac{(-1)^{s}}{s!}\left(\hat{a}^{\dag}(t)\right)^{m+s}\left(\hat{a}(t)\right)^{n+s}.
\end{align}
%
\section{}\label{C}
We have
\begin{equation}\label{C1}
  I_{\hat{B}}=\Tr_{R}\left\{e^{i\,\lambda\,\hat{B}^{\dag}(t)}e^{i\,\bar{\lambda}\,\hat{B}(t)}{\hat{\rho}}_{R}(0)\right\},
\end{equation}
where
\begin{align}\label{C2}
& \hat{B}(t)=\sum_{j}\left(M_{j}(t)\hat{b}_{j}(0)+2\,i\,\phi\,N_{j}(t)\hat{b}^{\dag}_{j}(0)\right),\nn\\
& \hat{B}^{\dag}(t)=\sum_{j}\left(\bar{M}_{j}(t)\hat{b}^{\dag}_{j}(0)-2\,i\,\bar{\phi}\,\bar{N}_{j}(t)\hat{b}_{j}(0)\right),\nn\\
& \hat{\rho}_{R}(t)=\frac{1}{Z_{R}}\prod_{j}\,e^{-\beta\hbar\omega_{j}\hat{b}^{\dag}_{j}\hat{b}_{j}}.
\end{align}
By inserting Eqs. (\ref{C2}) into Eq. (\ref{C1}) we obtain
\begin{equation}\label{C.1}
  I_{\hat{B}}=\prod_{k}\Tr_{k}\left\{e^{\left(i\,\lambda\bar{M}_{k}(t)\hat{b}^{\dag}_{k}(0)+2\,\lambda\,\bar{\phi}\,\bar{N}_{k}(t)\hat{b}_{k}(0)\right)}
  e^{\left(i\,\bar{\lambda}M_{k}(t)\hat{b}_{k}(0)-2\,\bar{\lambda}\,\phi\,N_{k}(t)\hat{b}^{\dag}_{k}(0)\right)}
  \frac{e^{-\beta\hbar\omega_{k}\hat{b}^{\dag}_{k}\hat{b}_{k}}}{Z_{k}}\right\},
\end{equation}
where $\Tr_{k}$ means taking trace over the Hilbert space of the $k$th oscillator in the environment and $Z_{k}$ is the corresponding partition function.
By using the following equations
\begin{align}
& e^{\mu\hat{b}_{k}(0)+\nu\hat{b}^{\dag}_{k}(0)}=e^{\mu\hat{b}_{k}(0)}e^{\nu\hat{b}^{\dag}_{k}(0)}e^{-\frac{1}{2}\mu\,\nu},\nn\\
& e^{\mu\hat{b}_{k}(0)}e^{\nu\hat{b}^{\dag}_{k}(0)}=e^{\nu\hat{b}^{\dag}_{k}(0)}e^{\mu\hat{b}_{k}(0)}e^{\mu\nu},
\end{align}
we have
\begin{align}\label{C3}
I_{\hat{B}}=& \prod_{k}\frac{1}{Z_{k}}\sum_{n_{k}=0}^{\infty}e^{-\beta\hbar\omega_{k}n_k}
\sum_{m=0}^{n_{k}}\frac{(-1)^{2}|\lambda|^{2m}|V_{k}(t)|^{2m}}{(m!)^{2}}\frac{n_{k}!}{(n_{k}-m)!}e^{i(\lambda^2\,\bar{\phi}\bar{N_k}(t) \bar{M_k}(t)-\bar{\lambda}^2 \phi N_k(t) M_k(t))-4|\lambda|^{2}|\phi|^{2}|N_{k}(t)|^{2}},\nn\\
=& e^{\sum\limits_{k}\Big[-\lambda\,\bar{\lambda}\Big(\frac{|V_{k}(t)|^{2}}{e^{\beta\hbar\,\omega_{k}}-1}+4|\phi|^2\,|N_k (t)|^2\Big)+i(\lambda^2\,\bar{\phi}\bar{N_k}(t) \bar{M_k}(t)-\bar{\lambda}^2 \phi N_k(t) M_k(t))\Big]}=e^{\vartheta[\lambda,\bar{\lambda},t]}.
\end{align}
%
\section{}\label{D}
We have
\beq
\hat{C}(t)=\alpha_{1}(t)\hat{a}(0)-2\,i\,\phi\,\alpha_{2}(t)\hat{a}^{\dag}(0),
\eeq
therefore,
\begin{align}\label{D0}
I_{\hat{C}}&=\Tr_{S}\left\{e^{\lambda\left(\bar{\alpha}_{1}(t)\hat{a}^{\dag}(0)+2\,i\,\bar{\phi}\,\alpha_{2}(t)\hat{a}(0)\right)}
e^{-\bar{\lambda}\left(\alpha_{1}(t)\hat{a}(0)-2\,i\,\phi\,\alpha_{2}(t)\hat{a}^{\dag}(0)\right)}\,\hat{\rho}_{S}(0)\right\},\nn\\
&=\Tr_{S}\left\{e^{\lambda\,\bar{\alpha}_{1}(t)\hat{a}^{\dag}(0)}e^{2\,i\,\lambda\,\bar{\phi}\alpha_{2}(t)\hat{a}(0)}
e^{i\,\lambda^{2}\bar{\phi}\,\bar{\alpha}_{1}(t)\alpha_{2}(t)}e^{-\bar{\lambda}\alpha_{1}(t)\hat{a}(0)}
e^{2\,i\,\bar{\lambda}\phi\,\alpha_{1}(t)\hat{a}^{\dag}(0)}e^{i\,\bar{\lambda}^{2}\phi\,\alpha_{1}(t)\alpha_{2}(t)}\,\hat{\rho}_{S}(0)\right\},\nn\\
 &= e^{\left(i\lambda^{2}\bar{\phi}\,\bar{\alpha}_{1}(t)-i\,\bar{\lambda}^{2}\phi\,\alpha_{1}(t)
  -4\,\lambda\bar{\lambda}|\phi|^{2}\alpha_{2}(t)\right)\alpha_{2}(t)}
  \Tr_{S}\left\{e^{\sigma\hat{a}^{\dag}}e^{-\bar{\sigma}\hat{a}}\hat{\rho}_{S}(0)\right\},
\end{align}
where
\begin{align}
\sigma=&\lambda\bar{\alpha}_{1}(t)+2\,i\,\bar{\lambda}\phi\alpha_{2}(t),\nn\\
\bar{\sigma}=&\bar{\lambda}\alpha_{1}(t)-2\,i\,\lambda\bar{\phi}\bar{\alpha}_{2}(t).
\end{align}
The normal and anti-normal characteristic functions in quantum optics are respectively defined by
\begin{align}\label{D.2}
  \Tr_{S}\left\{e^{\sigma\hat{a}^{\dag}}e^{-\bar{\sigma}\hat{a}}\hat{\rho}_{S}(0)\right\}&=C_{N}(\sigma),\nn\\
  \Tr_{S}\left\{e^{-\bar{\sigma}\hat{a}}e^{\sigma\hat{a}^\dag}\hat{\rho}_{S}(0)\right\}&=C_{A}(\sigma),
\end{align}
and are related as $C_{N}(\sigma)=e^{\left|\sigma\right|^{2}}\,C_{A}(\sigma)$. Let us assume that the initial state of the system is a coherent state
\begin{equation}
\hat{\rho}_{S}(0)=|\,\gamma\,\rangle\langle\,\gamma\,|,
\end{equation}
then
\begin{equation}\label{D.3}
 C_{A}(\sigma)=\int\,d^{2}\alpha\,Q(\alpha)e^{\sigma\,\bar{\alpha}-\bar{\sigma}\,\alpha},
\end{equation}
where $Q(\alpha)$ is the Husimi distribution function
\begin{equation}\label{D.4}
  Q(\alpha)=\frac{\langle\,\alpha\,|\hat{\rho}_{S}(0)|\,\alpha\,\rangle}{\pi}=\frac{e^{-|\alpha-\gamma|^{2}}}{\pi}.
\end{equation}
By inserting Eq. (\ref{D.4}) into Eq. (\ref{D.3}) we have
\begin{align}\label{D.6}
 C_{A}(\sigma)&=\frac{1}{2\pi}\int\,d^{2}\alpha\,e^{-|\alpha-\gamma|^{2}}e^{\sigma\,\bar{\alpha}-\bar{\sigma}\,\alpha}\nn\\
  &=\frac{1}{2\pi}\int\,d^{2}\alpha\,e^{-|\alpha|^{2}-|\gamma|^{2}+\left(\gamma+\sigma\right)\bar{\alpha}+\left(\bar{\gamma}+\bar{\sigma}\right)\alpha},\nn\\
  &=\frac{e^{-|\gamma|^{2}}}{2\pi}\int\,d\,q\,e^{-\frac{q^{2}}{2}+\frac{1}{\sqrt{2}}\left[(\gamma+\bar{\gamma})+(\sigma-\bar{\sigma})\right]q}
\int\,d\,p\,e^{-\frac{p^{2}}{2}+\frac{-i}{\sqrt{2}}\left[(\gamma-\bar{\gamma})+(\sigma+\bar{\sigma})\right]p},\nn\\
  &=e^{-|\sigma|^{2}}e^{\bar{\gamma}\sigma-\gamma\bar{\sigma}},
\end{align}
therefore,
\begin{align}\label{D.8}
  \Tr_{S}\left\{e^{\lambda\hat{C}^{\dag}}e^{-\bar{\lambda}\hat{C}}\hat{\rho}_{S}(0)\right\}=&
  e^{\left(i\lambda^{2}\bar{\phi}\,\bar{\alpha}_{1}(t)+i\,\bar{\lambda}^{2}\phi\,\alpha_{1}(t)
  -4\,\lambda\bar{\lambda}|\phi|^{2}\alpha_{2}(t)-2\,i\,\bar{\lambda}^{2}\phi\alpha_{1}(t)\right)\alpha_{2}(t)}\nn \\
  & \times e^{\bar{\gamma}\left(\lambda\bar{\alpha}_{1}(t)+2\,i\,\bar{\lambda}\phi\alpha_{2}(t)\right)}
  e^{-\gamma\left(\bar{\lambda}\alpha_{1}(t)-2\,i\,\lambda\bar{\phi}\alpha_{2}(t)\right)}.
\end{align}
%
\section{}\label{E}
To calculate $I$, we define on the complex plane
\begin{align}
& \lambda=u\,+i\,v,\,\,\,\,\,\,\,\,\,\,\,\,\,\,\bar{\lambda}=u\,-i\,v,\nn\\
& \partial_{\lambda}=\frac{1}{2}\partial_{u}+\frac{1}{2\,i}\partial_{v},\,\,\,\,\,\,\,\,\,\,\,\partial_{\bar{\lambda}}=\frac{1}{2}\partial_{u}-\frac{1}{2\,i}\partial_{v},
\end{align}
then
\begin{equation}
\partial_{\lambda}\partial_{\bar{\lambda}}=\frac{1}{4}\left(\partial_{u}^{2}+\partial_{v}^{2}\right)=\frac{1}{4}\nabla^{2}_{(u,v)}.
\end{equation}
By definition we have
\begin{equation}\label{E.1}
I(u,v)=e^{-\eta(t)\left(u^{2}\,+v^{2}\right)+\left(u\,+i\,v\right)\bar{Z}-\left(u\,-i\,v\right)Z},
\end{equation}
using the Fourier transform
\begin{equation}\label{E.2}
  I(u,v)=\int\int\,d^{2}\vec{k}\,e^{\left(i\,k_{u}\,u\,+\,i\,k_{v}\,v\right)}\tilde{I}(k_{u},k_{v}),
\end{equation}
one finds
\begin{align}\label{E.3}
  e^{\partial_{\lambda}\partial_{\bar{\lambda}}}I(\lambda,\bar{\lambda})&=e^{\frac{1}{4}\nabla^{2}_{(u,v)}}\,I(u,v),\nn \\
  &=\int\int\,d\,k_{u}\,d\,k_{v}\,e^{-\frac{1}{4}\left(k_{u}^{2}+k_{v}^{2}\right)+i\,k_{u}\,u\,+\,i\,k_{v}\,v}\tilde{I}(k_{u},k_{v}),\nn \\
  &=g(u,v).
\end{align}
Therefore,
\begin{eqnarray}\label{E.4}
  P_{n}(t)|_{\phi=0} &=& \frac{(-1)^n}{n!}\Big(\frac{\partial}{\partial\lambda}\frac{\partial}{\partial\bar{\lambda}}\Big)^{n}\,{e^{\partial_{\lambda}\partial_{\bar{\lambda}}}}
  I(\lambda,\bar{\lambda})\big|_{\lambda=\bar{\lambda}=0}, \nn\\
  &=& \frac{1}{(-4)^{n}n!}(\nabla^{2})^{n}g(u,v)\big|_{u=v=0}.
\end{eqnarray}
Now from the inverse Fourier transform
\begin{align}\label{E.5}
  \tilde{I}(k_{u},k_{v})&=\frac{1}{(2\,\pi)^2}\int\int\,d\,u\,d\,v\,e^{-i\left(k_{u}\,u\,+\,k_{v}\,v\right)}I(u,v),\nn\\
  &=\frac{1}{(2\,\pi)^2}\int\,d\,u\,e^{-\eta(t)u^{2}}e^{\left((\bar{Z}-Z)-i\,k_{u}\right)u}\int\,d\,v\,e^{-\eta(t)v^{2}}e^{\left(i(\bar{Z}+Z)-i\,k_{v}\right)v},
\end{align}
we have
\begin{equation}\label{E.6}
 \tilde{I}(k_{u},k_{v})=\frac{1}{4\pi\eta(t)}e^{\left(\frac{-Z\bar{Z}}{\eta(t)}\right)}e^{-\left(\frac{k_{u}^{2}}{4\eta(t)}+\frac{i(\bar{Z}-Z)k_{u}}{2\eta(t)}\right)}
e^{-\left(\frac{k_{v}^{2}}{4\eta(t)}-\frac{(\bar{Z}+Z)k_{v}}{2\eta(t)}\right)},
\end{equation}
leading to
\begin{align}\label{E.7}
  g(u,v)=& \left[-\frac{e^{\left(\frac{Z\bar{Z}}{-\eta(t)}\right)}}{4\pi\eta(t)}
  \int d\,k_{u}\,e^{\left(\frac{1}{-4\eta(t)}-\frac{1}{4}\right)k_{u}^{2}}e^{\frac{i}{-2\eta(t)}\left(\bar{Z}-Z-2\eta(t)\,u\right)k_{u}}
  \int d\,k_{v}\,e^{\left(\frac{1}{-4\eta(t)}-\frac{1}{4}\right)k_{v}^{2}}e^{-\frac{1}{2\eta(t)}\left(\bar{Z}+Z+2\,i\,\eta(t)\,v\right)k_{v}}\right],\nn\\
  =& -\frac{1}{-\eta(t)-1}e^{\frac{Z\bar{Z}}{-\eta(t)-1}}\left[e^{-\frac{\eta(t)}{\eta(t)+1}\left(u^{2}+v^{2}\right)}e^{-\frac{\bar{Z}-Z}{-\eta(t)-1}u}
 e^{i\frac{\bar{Z}+Z}{-\eta(t)-1}v}\right].
\end{align}
Therefore,
\begin{equation}\label{E.9}
  P_{n}(t)|_{\phi=0}=
  \frac{e^{-\frac{|Z|^2}{1+\eta(t)}}}{(-4)^{n}n!(1+\eta(t))}
  (\nabla^{2}_{(u,v)})^{n}e^{\frac{-\eta(t)u^{2}-\eta(t)v^{2}+(\bar{Z}-Z)u\,-\,i(\bar{Z}+Z)v}{1+\eta(t)}}\Big |_{u=v=0}.
\end{equation}
Now using the following identities
\begin{align}\label{E.10}
& \nabla^{2}_{(u,v)}=\partial_{u}^{2}+\partial_{u}^{2},\nn\\
& \left[\partial_{u},\partial_{v}\right]=0,\nn\\
&  \left(\partial_{u}^{2}+\partial_{u}^{2}\right)^{n}e^{\frac{-\eta(t)u^{2}+\left(\bar{Z}-Z\right)u}{1+\eta(t)}}e^{\frac{-\eta(t)v^{2}
  -i\left(\bar{Z}+Z\right)v}{1+\eta(t)}}
  =\sum_{k=0}^{\infty}\binom{n}{k}\left(\partial_{u}^{2}\right)^{n-k}e^{\frac{-\eta(t)u^{2}+\left(\bar{Z}-Z\right)u}{1+\eta(t)}}
  \left(\partial_{v}^{2}\right)^{k}e^{\frac{-\eta(t)v^{2}-i\left(\bar{Z}+Z\right)v}{1+\eta(t)}},\nn\\
& e^{-l^{2}+2\,x\,l}=\sum_{n=0}^{\infty}H_{n}(x)\frac{l^{n}}{n!},\nn\\
& \partial_{l}^{n}\left(e^{-l^{2}+2\,x\,l}\right)|_{l=0}=H_{n}(x),\nn\\
\end{align}
and the definitions
\begin{align}
& -\frac{\eta(t)u^{2}}{\eta(t)+1}=-l^{2},\,\,\,\,\,\,\,\,\,\,\,\,\,\,u=\sqrt{\frac{\eta(t)+1}{\eta(t)}}l,\nn\\
& x=\frac{\bar{Z}-Z}{2\sqrt{\eta(t)(\eta(t)+1)}},\nn\\
& -\frac{\eta(t)v^{2}}{\eta(t)+1}=-{l'}^{2},\,\,\,\,\,\,\,\,\,\,\,\,\,\,v=\sqrt{\frac{\eta(t)+1}{\eta(t)}}l',\nn\\
& x'=-i\frac{Z+\bar{Z}}{2\sqrt{\eta(t)(\eta(t)+1)}},
\end{align}
we will find
\begin{align}\label{E.11}
& e^{(\frac{-\eta(t)u^{2}+(\bar{Z}-Z)u}{1+\eta(t)})}=\sum_{p=0}^{\infty}
H_{p}\Big(\frac{\bar{Z}-Z}{2\sqrt{\eta(t)(\eta(t)+1)}}\Big)\frac{\bigg(\sqrt{\frac{\eta(t)}{\eta(t)+1}}\,u\bigg)^{p}}{p!},\nn\\
& e^{(\frac{\eta(t)v^{2}-i(\bar{Z}+Z)v}{1-\eta(t)})}=\sum_{q=0}^{\infty}
H_{q}\Big(\frac{-i(Z+\bar{Z})}{2\sqrt{\eta(t)(\eta(t)+1)}}\Big)\frac{\bigg(\sqrt{\frac{\eta(t)}{\eta(t)+1}}\,v\bigg)^{q}}{q!},
\end{align}
so
\begin{align}\label{E.13}
& \nabla^{2\,n}\left(e^{\frac{-\eta(t)u^{2}+\left(\bar{Z}-Z\right)u}{1+\eta(t)}}e^{\frac{-\eta(t)v^{2}-i\left(\bar{Z}+Z\right)v}
  {1+\eta(t)}}\right)\Big |_{u=v=0}=\nn\\
&  \sum_{k=0}^{n}\binom{n}{k}
\left[\partial_{u}^{2(n-k)}\sum_{p=0}^{\infty}H_{p}\left(\frac{\bar{Z}-Z}{2\sqrt{\eta(t)(\eta(t)+1)}}\right)
\left(\sqrt{\frac{\eta(t)}{\eta(t)+1}}\right)^{p}\frac{u^{p}}{p!}\right]_{u=0}\nn\\
  &\,\,\,\,\,\,\,\,\,\,\,\,\,\,\,\,\,\times\left[\partial_{v}^{2\,k}\sum_{q=0}^{\infty}H_{q}\left(\frac{-i(Z+\bar{Z})}{2\sqrt{\eta(t)(\eta(t)+1)}}\right)
\left(\sqrt{\frac{\eta(t)}{\eta(t)+1}}\right)^{q}\frac{v^{q}}{q!}\right]_{v=0},\nn\\
& =\sum_{k=0}^{n}\binom{n}{k}\left(\frac{\eta(t)}{\eta(t)+1}\right)^{n}H_{2\,n-2\,k}\left(\frac{\bar{Z}-Z}{2\sqrt{\eta(t)(\eta(t)+1)}}\right)
H_{2\,k}\left(\frac{-i(Z+\bar{Z})}{2\sqrt{\eta(t)(\eta(t)+1)}}\right).
\end{align}
Finally, using the identity
\begin{equation}\label{E.16}
\sum_{k=0}^{n}\binom{n}{k}H_{2\,n-2\,k}(x) H_{2\,k}(y)=(-4)^{n}n!\,L_{n}(x^{2}+y^{2}),
\end{equation}
we deduce that
\begin{align}\label{E.17}
 P_{n}(t)|_{\phi=0}&=\frac{e^{-\frac{|Z|^2}{1+\eta(t)}}}{(-4)^{n}n!(1+\eta(t))}\left(\frac{\eta(t)}{\eta(t)+1}\right)^{n}
 \left[(-4)^{n}n!\,L_{n}\left(\frac{(Z-\bar{Z})^{2}-(Z+\bar{Z})^{2}}{4\eta(t)(\eta(t)+1)}\right)\right]\nn \\
  &=\frac{e^{-\frac{|Z|^2}{1+\eta(t)}}}{(1+\eta(t))}\left(\frac{\eta(t)}{\eta(t)+1}\right)^{n}
 L_{n}\left(\frac{-|Z|^{2}}{\eta(t)(1+\eta(t))}\right),
\end{align}
where $L_n [x]$ is a Laguerre polynomial of degree $n$.


\begin{thebibliography}{00}




\bibitem{F.1} H.-P. Breuer, F. Petruccione, \emph{The Theory of Open Quantum Systems} (Oxford University Press, Oxford,
2002)
\bibitem{F.2} A. Lampo, \emph{Quantum Brownian motion revisited: extensions and applications}. Doctorial thesis, (Universitat
Politècnica de Catalunya, Catalunya) (2018)
\bibitem{F.3} U. Weiss, \emph{Quantum Dissipative Systems}, 2nd edn. (World Scientific, Singapore, 1999)
\bibitem{F.4} A.O. Caldeira, \emph{An Introduction to Macroscopic Quantum Phenomena and Quantum Dissipation} (Cambridge
University Press, Cambridge, 2014)
\bibitem{F.5} J. Schwinger, \emph{Brownian Motion of a Quantum Oscillator}, J. Math. Phys. 2, 407 (1961)
\bibitem{F.6} R. P. Feynman, F. L. J. Vernon, \emph{The theory of a general quantum system interacting with a linear dissipative system}, Ann. Phys. (N.Y.) 24, 118 (1963)
\bibitem{F.7} A.O. Caldeira, A. J. Leggett, \emph{Path integral approach to quantum Brownian motion}, Physica A 121, 587 (1983)
\bibitem{F.8} H. Grabert, P. Schramm, G.-L. Ingold, \emph{Quantum Brownian motion: The functional integral approach}, Phys. Rep. 168, 115 (1988)
\bibitem{F.9} M. Carlesso, A. Bassi, \emph{Adjoint master equation for quantum Brownian motion}, Phys. Rev. A 95, 052119 (2017)
%
\bibitem{G.1} J. Gemmer, M. Michel, G. Mahler,\emph{ Quantum Thermodynamics}, 2nd edn. (Springer, Berlin, 2009)
\bibitem{G.2} S. Gasparinetti, P. Solinas, A. Braggio, M. Sassetti, \emph{Heat-exchange statistics in driven open quantum systems}, New J. Phys. 16, 115001 (2014)
\bibitem{G.3} M. Carrega, P. Solinas, A. Braggio, M. Sassetti, U. Weiss, \emph{Functional integral approach to time-dependent heat exchange in open quantum systems: general method and applications}, New J. Phys. 17, 045030 (2015)
\bibitem{G.4} W. Dou, M. A. Ochoa, A. Nitzan, J. E. Subotnik, \emph{Universal approach to quantum thermodynamics in the strong coupling regime}, Phys. Rev. B 98, 134306 (2018)
\bibitem{G.5} R. S. Whitney,\emph{ Non-Markovian quantum thermodynamics: Laws and fluctuation theorems}, Phys. Rev. B 98, 085415 (2018)
\bibitem{G.6} K. Funo, H. T. Quan, \emph{Path Integral Approach to Quantum Thermodynamics}, Phys. Rev. Lett. 121, 040602 (2018)
\bibitem{G.7} M. Perarnau-Llobet, H. Wilming, A. Riera, R. Gallego, J. Eisert, \emph{Strong Coupling Corrections in Quantum Thermodynamics}, Phys. Rev. Lett. 120, 120602 (2018)
\bibitem{G.8} M. A. Ochoa, N. Zimbovskaya, A. Nitzan,\emph{ Quantum thermodynamics for driven dissipative bosonic systems}, Phys. Rev. B 97, 085434 (2018)
\bibitem{G.9} P. Haughian, M. Esposito, T. L. Schmidt, \emph{Quantum thermodynamics of the resonant-level model with driven system-bath coupling}, Phys. Rev. B 97, 085435 (2018)
\bibitem{G.10} J. Lekscha, H. Wilming, J. Eisert, R. Gallego,\emph{ Quantum thermodynamics with local control}, Phys. Rev. E 97, 022142 (2018)
%
\bibitem{Mandal} S. Mandal, \emph{SQUEEZING, HIGHER-ORDER SQUEEZING, PHOTON-BUNCHING AND PHOTON-ANTIBUNCHING IN A QUADRATIC HAMILTONIAN}, Mod. Phys. Lett. B 16, 963 (2002)
\bibitem{Tsai} S. -W. Tsai A. F. R. de Toledo Piza, \emph{Kinetics of photon correlation functions under the time-dependent quadratic Hamiltonian}, Phys. Rev. A 53, 3683 (1996)
\bibitem{Piza} A. F. R. de Toledo Piza, \emph{Classical equations for quantum squeezing and coherent pumping by the time-dependent quadratic Hamiltonian}, Phys. Rev. A 51, 1612 (1995)
%
\bibitem{H1} M. Tokieda and K. Hagino, \emph{A new approach for open quantum systems based on a phonon number representation of a harmonic oscillator bath}, Ann. Phys. 412, 168005 (2020)
\bibitem{H2} P. A. Golovinski, \emph{Dynamics of driven Brownian inverted oscillator}, Phys. Lett. A 384, 126203 (2020)
\bibitem{H3} V. A. Tomilin and L. V. I\`{L}ichov, \emph{Solvable model of quantum-optical feedback}, Phys. Lett. A 384, 126718 (2020)
\bibitem{H4} A. S. Pereira and A. S. Lemos, \emph{Time-dependent coherent squeezed states in a nonunitary approach}, Phys. Lett. A 405, 127428 (2021)
\bibitem{H5} V. V. Dodonov, \emph{Invariant Quantum States of Quadratic Hamiltonians}, Entropy 23, 634 (2021)
\bibitem{H6} Tian Qiu and Hai-Tao Quan, \emph{Quantum corrections to the entropy in a driven quantum Brownian motion model}, Commun. Theor. Phys. 73, 095602 (2021)
%
\bibitem{Basaeia} B. Basaeia and C. A. Bonato, \emph{Squeezing in Systems Described by Quartic Hamiltonians: Normal Ordering Technique}, Il Nuvo Cimento, Vol. 107B, No.9 (1992)
\bibitem{Toledo} A. F. R. de Toledo Piza, \emph{Classical equations for quantum squeezing and coherent pumping by the time-dependent quadratic Hamiltonian}, Phys. Rev. A51, 1612 (1995)
\bibitem{Ben} Y. Ben-Aryeh and H. Zoubi, \emph{The time development operators for Wigner functions of harmonic oscillators with quadratic Hamiltonians}, Quantum. Semiclass. Opt 8, 1097 (1996)
\bibitem{Choi} J. R. Choi, \emph{Dynamics of SU(1, 1) coherent states for the time-dependent quadratic Hamiltonian system}, Opt. Commu. 282, 3720 (2009)
\bibitem{Wang} S. Wang, H. -C. Yuan and H. -Y. Fan, \emph{FRESNEL OPERATOR, SQUEEZED STATE AND WIGNER FUNCTION FOR CALDIROLA–KANAI HAMILTONIAN}, Mod. Phys. Letts. A Vol. 26, No. 19, 1433 (2011)
\bibitem{Gilmore} W. -M. Zhang, D. H. Feng and R. Gilmore, \emph{Coherent states: Theory and some applications},  Rev. Mod. Phys. Vol. 62, No. 4, 867 (1990)
%
\bibitem{Langevin} W. T. Coffey, Yu. P. Kalmykov and J. T. Waldron, \emph{The Langevin Equation}, (World Scientific, 2004)
%

\bibitem{Louisell} W. H. Louisell, Quantum Statistical Properties of Radiation (Wiley, New York, 1975)

\end{thebibliography}
\end{document}